\documentclass[9pt,twocolumn,twoside]{osajnl}

\usepackage{color}
\def\0#1{\xout{#1}}
\def\1#1{{\color{red}#1}}
\def\2#1{{\color{green}#1}}
\def\3#1{{\color{blue}#1}}
\def\4#1{{\color{yellow}#1}}
\usepackage{multirow}

\journal{josab} 

\setboolean{shortarticle}{false} 

\title{Advanced phase retrieval for dispersion scan: a comparative study}

\author[1*]{Esmerando Escoto}
\author[1]{Tamas Nagy}
\author[2]{Ayhan Tajalli}
\author[1]{G{\"u}nter Steinmeyer}

\affil[1]{Max-Born-Institute for Nonlinear Optics and Short Pulse Spectroscopy, Max-Born-Stra{\ss}e 2a, 12489 Berlin, Germany}
\affil[2]{Institute of Quantum Optics, Leibniz Universit{\"a}t Hannover, Welfengarten 1, D-30167 Hannover, Germany}

\affil[*]{Corresponding author: escoto@mbi-berlin.de}

\ociscodes{(320.7100) Ultrafast measurements; (190.7110) Ultrafast nonlinear optics.}

\doi{}

\begin{abstract}
Dispersion scan is a self-referenced measurement technique for ultrashort pulses. Similar to frequency-resolved optical gating, the dispersion scan technique records the dependence of nonlinearly generated spectra as a function of a parameter. For the two mentioned techniques, these parameters are the delay and the dispersion, respectively. While dispersion scan seems to offer a number of potential advantages over other characterization methods, in particular for measuring few-cycle pulses, retrieval of the spectral phase from the measured traces has so far mostly relied on the Nelder-Mead algorithm, which has a tendency of stagnation in a local minimum and may produce ghost satellites in the retrieval of pulses with complex spectra. We evaluate three different strategies to overcome these retrieval problems, namely regularization, use of a generalized-projections algorithm, and an evolutionary retrieval algorithm. While all these measures are found to improve the precision and convergence of dispersion scan retrieval, differential evolution is found to provide best performance, enabling the near-perfect retrieval of the phase of complex supercontinuum pulses within less than ten seconds, even in the presence of strong detection noise and limited phase-matching bandwidth of the nonlinear process.
\end{abstract}

\setboolean{displaycopyright}{false}

\begin{document}

\maketitle

\pagestyle{plain}
\thispagestyle{plain}

\section{Introduction}

In the past 20 years, a number of self-referenced measurement techniques have emerged that enable complete reconstruction of the amplitude and  relative phase structure of an ultrashort laser pulse. Frequency-resolved optical gating (FROG, \cite{Trebino:93,FROGBook}) is the pioneer among these techniques and relies on measuring spectrally resolved autocorrelation traces. Conceptually related methods include the sonogram method \cite{Reid99,sonogram}, multiphoton intrapulse interference phase scan (MIIPS, \cite{MIIPS2,MIIPS}), dispersion scan (d-scan, \cite{Miranda:12a,Miranda:12b}), and chirp scan \cite{chirpscan}. While the latter two related techniques are only a few years old, dispersion scan has recently proven a remarkable potential to accurately measure few-cycle pulses generated by compression of hollow-fiber supercontinua \cite{Timmers,Tajalli:16}. All these techniques effectively measure the dependence of the spectrum of a nonlinearly generated optical signal as a function of one particular parameter, e.g., the delay between two pulses in FROG. The dependence of the spectra upon variation of the parameter is then visualized as a two-dimensional image, such as the FROG trace. In all the above-mentioned methods it is straightforward to compute the respective two-dimensional trace for a known amplitude and phase structure. However, reconstruction of the pulse shape from a measured trace is an inverse problem, which requires an algorithmic approach. Independent of technique employed, the inverse problem in pulse retrieval is relatively easy to solve for pulses with time-bandwidth products close to the Fourier limit. Nevertheless, retrieval becomes more challenging with increasing complexity. Here, FROG certainly sets the standard, with a more than 20 year record of continuous development that has led to a number of highly sophisticated algorithms \cite{Fienup87,Trebino:93,DeLong:94a,DeLong:94b,Kane99,Kane08,Hause15}. However, all self-referenced pulse characterization techniques still face difficulties with correctly retrieving more complex pulses. In the presence of a well-characterized reference pulse, cross-correlation based methods certainly offer an alternative \cite{XFROG}, yet often enough such a reference pulse is not available.

FROG retrieval nearly exclusively relies on the generalized projections algorithm (GP, \cite{DeLong:94b,Kane08}), which uses a similar structure as the Gerchberg-Saxton algorithm \cite{Gerchberg} for two-dimensional phase retrieval. The GP approach tries to obtain best agreement of reconstructed and measured FROG trace while simultaneously enforcing a physical constraint that is dictated by the nonlinear process employed in the measurement. The data constraint is enforced by a projection in the frequency domain whereas the physical constraint is applied in the time domain. While GP has been applied to sonogram traces \cite{sonogram}, d-scan \cite{Miranda:12a,Miranda:12b} mostly relied only on the Nelder-Mead (NM) strategy \cite{Nelder01011965}. Only very recently, an improved retrieval strategy was reported that relied on GP \cite{arxiv}. Moreover, several other methods including bound optimization by quadratic approximation, the Broyden-Fletcher-Goldfarb-Shanno algorithm, and the Levenberg-Marquardt algorithm have been explored for retrieval in d-scan \cite{private,BOBYQA,BFGS,LMA1,LMA2}, and found to accelerate convergence relative to NM. Nelder-Mead is very generally applicable, and it reliably retrieves simply structured pulse shapes near the transform limit. However, we often observe early stagnation of the NM algorithm in the retrieval of moderately complex pulses from a compressed hollow-fiber supercontinuum \cite{hollowfiberdata}. A second frequently observed problem in the NM-based pulse retrieval is a build-up of artificial oscillations in the spectral phase.

In the following sections, we first discuss the dispersion scan method in detail along with the algorithm used to retrieve pulse characteristics from the d-scan traces. We then discuss several advanced concepts for pulse retrieval from d-scan traces, using either regularization \cite{Birkholz:15}, the GP, or an evolutionary algorithm \cite{diffevolution}. The main purpose of this comparison is to find a robust method that can automatically retrieve complex pulse shapes without user interaction and without going through a large number of initial random seeds. A secondary consideration is the computation time required for the retrieval. While we restrict ourselves to d-scan as it seems the technique which would benefit most from improved retrieval algorithms, we emphasize that some of the conclusions from this study appear rather universal, in particular when it comes to the reconstruction of rather complex pulses as they are generated in hollow-fiber compressors. We are confident that at least one of the retrieval strategies can also be beneficially used for FROG.

\section{The dispersion scan method}

\begin{figure}[htbp]
\centering
\includegraphics[width=.7\linewidth]{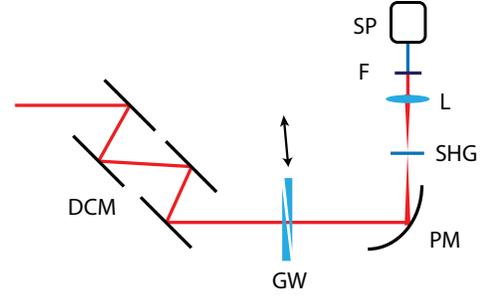}
\caption{Setup for measurement of dispersion scans. (DCM - double chirped mirrors, GW - glass wedges, PM - parabolic mirror, SHG - second harmonic generation crystal, L - lens, F - filter, SP - spectrometer)}
\label{fig:dscansetup}
\end{figure}

The dispersion scan (d-scan) technique is a method for measuring the temporal profile of ultrashort laser pulses, and it resembles FROG and sonogram in many aspects as it also relies on measuring nonlinearly generated spectra under variation of a parameter of the input pulse, which is the group delay dispersion in this case \cite{Miranda:12a,Miranda:12b}. The parameter variation is accomplished by a combination of chirped mirrors and a pair of adjustable glass wedges, recording the nonlinearly generated spectra  as a function of glass insertion, see Fig.~\ref{fig:dscansetup}. Conceptually, this is very similar to recording a FROG trace, yet replacing the autocorrelation delay in FROG by the varying group delay dispersion (GDD) induced by the wedges. D-scan is typically used with a set of chirped mirrors to provide a fixed amount of negative dispersion, i.e., the spectra are recorded around the maximum compression of the pulse. In the following, we use a representation of the d-scan trace with a wavelength axis $\lambda$ and a glass insertion axis $z$. For $z=0$, the chirped mirrors exactly cancel out the dispersion of the wedges. As a nonlinearity, one typically employs second-harmonic generation (SHG) in d-scan, but it is also possible to use third-order nonlinear processes \cite{Hoffmann:14,Tajalli:16,Canhota:17}. Here we restrict ourselves to SHG, but we are confident that our conclusions also hold for third-order d-scan.

As shown in Fig.~\ref{fig:dscansetup}, d-scan relies on a single-beam geometry and is therefore very simple to set up experimentally. Moreover, avoiding the crossed-beam correlator geometry utilized in most other pulse characterization schemes, one can employ hard focusing into the nonlinear medium, which enables the use of weak third-order nonlinearities \cite{Tajalli:16} or the measurement of faint pulses with SHG.

\begin{figure}[htbp]
\centering
\includegraphics[width=0.7 \linewidth]{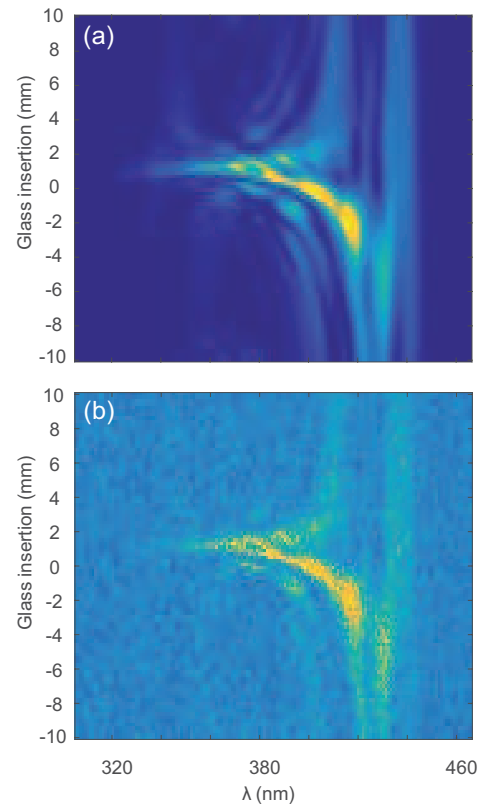}
\caption{D-scan traces with the minimum and maximum noise levels used in this study: (a) 0.1\%, and (b) 10\%. Noise has been modeled as additive Gaussian white noise.}
\label{fig:dscantraces}
\end{figure}

The two-dimensional traces from a d-scan measurement can be written as
\begin{equation}
\label{eq:dscantrace}
I_{\rm DS}(\omega,z) =  \left| \tilde{E}_{\rm sig}(\omega,z) \right|^2 =  \bigg\lvert \int  E_{\rm sig}(t,z)
\exp (-i\omega t) \mathrm{d}t \bigg\rvert ^2,
\end{equation}
where the signal field is formally defined as
\begin{equation}
\label{eq:signalfield}
E_{\rm sig}(t,z) = \left( \int \tilde{E}(\omega) \exp (izk(\omega) + i\omega t ) \mathrm{d}\omega \right)^2.
\end{equation}
Here $z$ is the thickness of the dispersive material in the beam path, and $k(\omega)$ accounts for dispersion of the material. In order to emphasize similarities and differences between FROG and d-scan, we use a FROG-like notation throughout. In particular, it is striking that the fundamental structure of the two-dimensional traces is identical when the signal field is written as the square of a chirped field in the time domain. At first sight, this may look more complicated than the simple product form typically encountered in FROG, e.g., $E_{\rm sig}(t,\tau)=E(t)E(t-\tau)$ for SHG FROG. Nevertheless, considering that determination of the chirp of $E(t)$ is ultimately the goal of phase retrieval, variation of the material insertion is much closer related to the chirp than is the delay variation in autocorrelation geometries.

In general, there exist two different strategies for pulse shape retrieval. In FROG, it is customary to simultaneously retrieve the spectral amplitude $|E(\omega)|$ and phase $\phi(\omega)$, which delivers the possibility for comparison with an independently measured spectrum of the fundamental. This test is known as the frequency marginal check \cite{DeLong96}. For increasing bandwidth of the pulses, it becomes increasingly difficult to fulfill the marginal check, and methods have been explored to correct, e.g., for insufficient phase-matching bandwidth \cite{Baltuska}, which can be accounted for by a response function $R(\omega)$. In the case of broadband spectra, therefore, one can resort to phase-only retrieval. Combining the retrieved phase with an independent measurement of $|E(\omega)|$ then gives access to the full complex field of the pulses. Depending on the knowledge of the exact phase-matching characteristics, phase-only retrieval still enables a basic marginal check, comparing the computed $R(\omega)$ from the crystal characteristics with the function reconstructed within the retrieval algorithm. It should further be noted that phase-only retrieval can certainly reduce phase-matching issues to some extent. However, the ultimate limitation of this correction is reached when the spectral response function reaches near-zero values within the bandwidth of the pulse.

As a testbed for the various algorithm discussed below, we generated a synthetic phase function combined with the electric field amplitude $| \tilde{E}(\omega) |$ found in the supercontinuum measurements \cite{Tajalli:16}. To get this amplitude, pulses of 35 fs from a Ti:sapphire-CPA system with pulse energy of 2 mJ were spectrally broadened in a 3-m long stretched hollow-core waveguide filled with argon gas at a pressure of 250 mbar to a Fourier limit of 5.5 fs. From this data set, we computed the SHG d-scan trace according to Eq.~(\ref{eq:dscantrace}), with the spectral window discretized to $N=158$ points and with $M=100$ glass steps of step size 0.2 mm. The spectral window is chosen wide enough to host the entire second harmonic spectrum, but it can certainly be made smaller as d-scan has been shown to tolerate some spectral clipping \cite{Miranda:12a,Miranda:12b}. We multiplied the resulting traces with a spectral efficiency curve $R(\omega)$ corresponding to the phase-matching efficiency of a 20 $\mu$m thick BBO crystal. To enable a test on the robustness of the retrieval methods, we simulated synthetic detection noise \cite{Fittinghoff95} as additional Gaussian white noise with standard deviation in the range from 0.1\% to 10\% of the peak value, see Fig.~\ref{fig:dscantraces}. The continuous functions $\tilde{E}(\omega)$ and $R(\omega)$ are discretized as $n$-element vectors $|E| = \{E_1, E_2, \dots E_n\}$ and $\phi = \{\varphi_1, \varphi_2, \dots \varphi_n\}$ for the modulus and phase of the field. The $n$ used in the algorithms is equal to 90, which covers the full fundamental spectrum. Here $\omega = \{\omega_1, \omega_2, \dots \omega_n\}$, and $E(\omega_j)=E_j \exp ( i \phi_j )$. $R$ is represented by $\{R_1, R_2, \dots R_N\}$ when needed. As some of the algorithms require several instances of $\phi$ and $|E|$ in parallel, it is important not confuse the different instances of $\phi^{(j)}$ and the components of $\phi$, i.e, $\varphi_j$, and also $N$ as the number of points in the d-scan trace spectra not with $n$ for the fundamental spectra in the following.

\section{Nelder-Mead phase retrieval}
\label{sec:NM}

With the exception of \cite{arxiv}, all published work on d-scan retrieval relied exclusively on the Nelder-Mead (NM) approach \cite{Miranda:12a,Miranda:12b}. NM is a multi-dimensional, heuristic search algorithm. It only requires direct evaluation of a multi-dimensional function without computing its derivatives. This creates flexibility for the function being optimized. Nevertheless, the adaptability of the NM algorithm comes at the price of a relatively slow convergence compared to gradient methods. Moreover, as all other local minimization strategies, the method may fail to converge to the global minimum.

From an algorithmic point of view, phase-only retrieval is equivalent to an $n$-dimensional minimization. If both spectral phases and amplitudes are to be recovered, which is possible when $R(\omega)$ is known or assumed constant, the number of dimensions doubles. As this substantially increases the complexity of the inverse problem, such full retrieval has so far been avoided for the d-scan method, and the modulus of the electric field $|\tilde{E}(\omega)|$ is derived from an independent measurement of the spectrum. In NM-based retrieval of d-scan traces, a simplex composed of $n+1$ vertices is initially created. Each vertex of the simplex consists of an $n$-element array $\phi^{(i)}$. Using Eq.~(\ref{eq:dscantrace}) for computation of the simulated field, the goodness of fit is then evaluated according to \cite{Miranda:12a}
\begin{equation}
\label{eq:fitness}
G = \sqrt{\frac{1}{N M} \sum_{i=1}^N \sum_{j=1}^M \left( I_{\rm DS}^{\rm (meas)}(\omega_i,z_j) - R_i I_{\rm DS}^{\rm (sim)} (\omega_i,z_j) \right) ^2 }
\end{equation}
and provides a criterion for the similarity between measured and simulated d-scan trace, $I_{\rm DS}^{\rm (meas)}(\omega_i,z_j)$ and $I_{\rm DS}^{\rm (sim)}(\omega_i,z_j)$, respectively. In d-scan, Eq.~(\ref{eq:fitness}) plays a similar role as the FROG error. The spectrometer response and the SHG  conversion efficiency are both accounted for by $R_i$, which is calculated and updated every iteration using
\begin{equation}
R_i = \frac{\sum_j I_{\rm DS}^{\rm (meas)}(\omega_i,z_j) I_{\rm DS}^{\rm (sim)}(\omega_i,z_j)}{\sum_j I_{\rm DS}^{\rm (sim)}(\omega_i,z_j)^2}.
\end{equation}

After initialization of the simplex with vertices $\phi^{(i)}$, where $i=1, \dots , n+1$, the $G$ value for each $\phi^{(i)}$ is computed, and the vertex with highest $G$ is subsequently replaced. As shown in detail in Algorithm \ref{alg:NM}, the algorithm then goes through three different strategies for the replacement. If neither of these is successful the entire simplex is shrunk towards its centroid point. Conventional values for the coefficients of these steps were used ($\alpha = 1$, $\gamma = 2$, $\rho = 0.5$, and $\sigma = 0.5$). The algorithm then repeats all steps until stagnation is observed, for which two stopping criteria are used. The convergence criterion stops the algorithm if the difference between the best and the worst vertices is less than a certain value. The algorithm is also stopped when it exceeds a maximum number of function evaluations. These criteria were used consistently for the different algorithms, but can certainly be adjusted for enforcing better convergence at the expense of a longer computation time.

\begin{algorithm}[bt]
\caption{Nelder-Mead algorithm for d-scan}\label{alg:NM}
\begin{algorithmic}[1]
\State $\phi^{(i)} \gets \phi^{(1)}+ \Gamma(i-1)\rm{rand}(n)$ \Comment{initial simplex, $\Gamma \ll 1$}
\State Evaluate all $G^{(i)}$
\While{criteria not met}
\State sort $\phi^{(i)}$ according to $G^{(i)}$ ($\phi^{(n+1)}$ being the worst)
\State  $\phi^{(o)} \gets (\sum_i^n \phi^{(i)})/n $ \Comment{centroid}
\State  $\phi^{(r)} \gets \phi^{(o)} + \alpha (\phi^{(o)} - \phi^{(n+1)}) $ \Comment{reflection}
\If{$G^{(1)} < G^{(r)} < G^{(n)}$}
\State $\phi^{(n+1)} \gets \phi^{(r)}$
\EndIf
\If{$G^{(r)} < G^{(1)}$}
\State  $\phi^{(e)} \gets \phi^{(o)} + \gamma (\phi^{(r)} - \phi^{(o)}) $  \Comment{expansion}
\If{$G^{(e)} < G^{(r)}$}
\State $\phi^{(n+1)} \gets \phi^{(e)}$
\Else
\State $\phi^{(n+1)} \gets \phi^{(r)}$
\EndIf
\EndIf
\If{$G^{(n+1)} < G^{(r)}$}
\State  $\phi^{(c)} \gets \phi^{(o)} + \rho (\phi^{(n+1)} - \phi^{(o)}) $  \Comment{contraction}
\Else
\State  $\phi^{(i)} \gets \phi^{(1)} + \sigma (\phi^{(i)} - \phi^{(1)}) $  \Comment{shrinking}
\EndIf
\EndWhile
\State \textbf{return} $\phi^{\rm (best)}$
\end{algorithmic}
\end{algorithm}

As the NM algorithm is fairly slow for high-dimensional optimization problems, a possible workaround is the use of a basis function for the spectral phase. For example, in \cite{Miranda:12a} a Fourier series was suggested for this purpose. While this measure certainly accelerates the convergence it also increases the tendency of local stagnation, which can possibly be overcome by repeated restarts with random seeds or by introducing a random noise. However, we found that neither of these measures can enforce global optimization \cite{Lewis2000191} in the retrieval of the rather complex d-scan traces in Fig.~\ref{fig:dscantraces}. Another optimization suggested in \cite{Miranda:12b} is the use of a sparse representation of the phase and increasing the resolution whenever the algorithm stagnates. This technique has been used consistently in this work to optimize the performance of NM.

In conventional NM, the initial simplex is formed from the guessed phase by changing one of the $n$ components to create each vertex, resulting in $n+1$ vertices that include the initial guess. However, this construction cannot be employed for sparse representation  since there are less components used than vertices. Instead, the other vertices are then formed by adding a small random noise to the guess phase, as shown in Algorithm \ref{alg:NM}. A second problem may arise in spectral regions with low spectral power where large variations of the retrieved phase may occur. Modifications and alternative algorithms are presented in the following chapters to address these issues.

\begin{figure*}[tb]
\centering
\includegraphics[width=\linewidth]{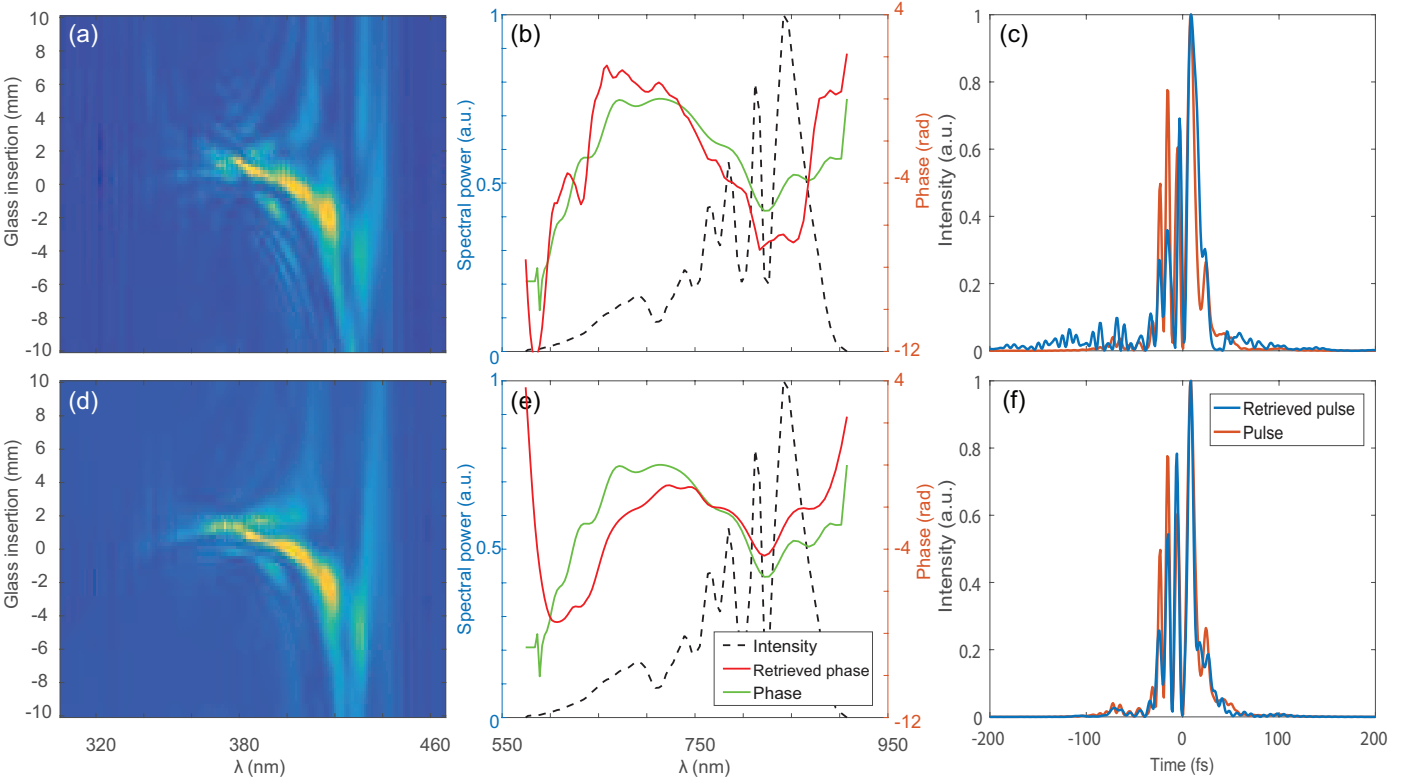}
\caption{ Retrieved d-scan trace and spectral phase, (a-c) without a penalty term, resulting to $G = 0.0999$ and (d-f) with a penalty term, resulting to $G = 0.0908$. These are the results for one trial using 10\% additive noise, as shown in Fig. \ref{fig:dscantraces}(f). }
\label{fig:regularized}
\end{figure*}

\section{Regularization}

Regularization is a technique used for inverse problems, such as phase retrieval, to avoid overfitting and to accelerate convergence by avoiding instabilities. Regularization has recently been used for a new variant of SPIDER, which is based on self-diffraction \cite{Birkholz:15}. Instabilities in the NM method  may arise due to a dense discretization of the continuous functions, e.g., when oversampling the electric field with more than one point per field cycle in the time domain. It is sometimes unavoidable to use dense spectral sampling, e.g., to accomodate the detailed spectral structure of a supercontinuum spectrum. Transforming back into the time domain, the fine spectral grid may then correspond to a picosecond temporal range, and even the best retrieval techniques encounter a challenge in correctly reconstructing the exact structure of numerous faint satellites. Suppression of rapidly oscillating spectral features therefore acts similar to confinement to a narrow interval in the time domain. Such confinement is exactly what regularization enforces. Otherwise, if the discretization is either locally or globally chosen too dense, one often observes the buildup of artificial oscillations in the retrieved spectral phase. An example for this undesired behavior is shown in Fig.~\ref{fig:regularized}(b), where phase oscillation artifacts appear during NM retrieval. In the time domain, these oscillations correspond to ghost satellite pulses appearing before or after the main pulse. To implement regularization for d-scan measurements, a penalty term, $P$, can be added to a loss function, which in this case is $G$ in Eq.~(\ref{eq:fitness}). The new function to be optimized then reads
\begin{equation}
G_R = G + \lambda^\ast P
\end{equation}
where $\lambda^\ast$ is a constant that determines the strength of regularization. For suppressing these artifacts, one can construct a penalty term as the sum of the modulus of some derivative of the phase. Numerically, one can implement this term by computing
\begin{equation}
P = \sqrt{\sum_{i=1}^{N-1} \left\lvert \varphi_i-\varphi_{i+1} \right\rvert ^2 }.
\end{equation}
Provided careful choice of the regularization parameter $\lambda^\ast$, regularization then prevents the buildup of extraneous phase oscillations, see Fig.~\ref{fig:regularized}(d). In particular, the fit in the central part of the spectrum is substantially improved by the regularization. In the time domain, a fairly large number of ghost pre-pulses at delays $<-50$\,fs are effectively suppressed, compare Figs.~\ref{fig:regularized}(c) and (f). However, retrieval of the exact structure of the four strongest sub-pulses is only marginally improved. Effectively, the penalty term therefore limits the temporal window of the reconstruction. Nevertheless, the fit is still far from perfect, and the convergence speed is still limited by the NM algorithm. While regularization may also slow down an algorithm, we typically observed a slight acceleration of the convergence in our numerical experiments, cf.~Table~\ref{tab:algo}.

We conclude that regularization certainly improves the retrieval capabilities in case of pulses with complex structure, but then it also limits the complexity due to the penalty term, and the parameter $\lambda^\ast$ has to be carefully adapted to the situation. Therefore regularization does not appear to be the ideal solution for the d-scan retrieval problem.

\section{Generalized Projections}

\begin{figure*}[tb]
\centering
\includegraphics[width=\linewidth]{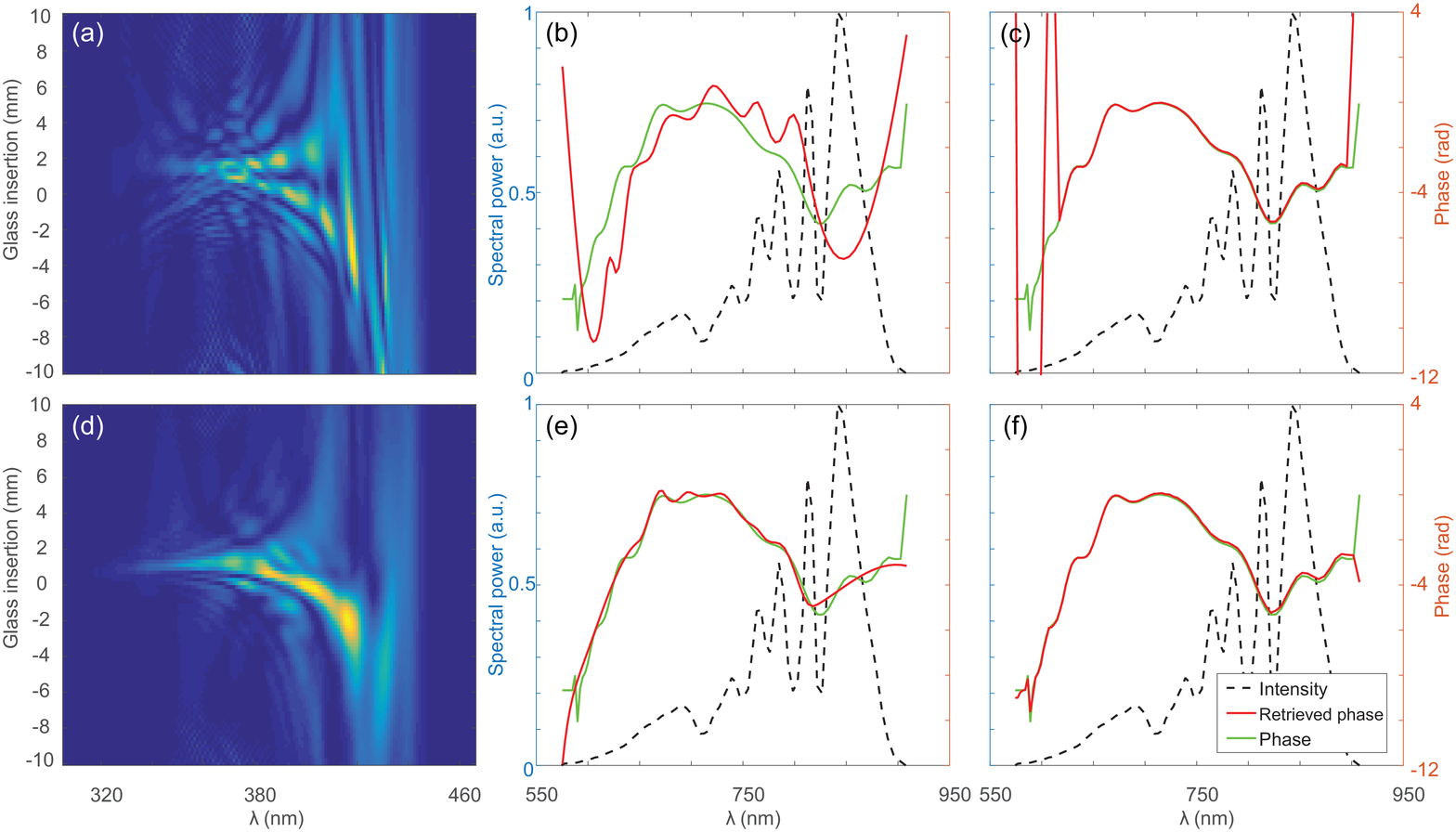}
\caption{Retrieved d-scan trace and spectral phase, (a-b) using the original algorithm, resulting to $G = 0.0882$, and (d-e) using the generalized projections structure, resulting to $G = 0.0259$. These are the results for one trial using 0.5\% additive noise. The results of letting the algorithm run for a much longer time for the original and the generalized projections algorithms are shown in (c) and (f) respectively. Despite this difference in retrieved phase, both retrieved traces (not shown) from (c) and (f) have $G = 0.0012$.}
\label{fig:GP}
\end{figure*}

As the second retrieval algorithm, we employed the generalized projections (GP) method, which is generally favored in FROG. GP takes advantage of the access to both time and frequency domains of the solution, where constraints can be applied in both representations. While early implementation of the GP method were also prone to occasional lockups, modern implementations have been found fast, robust, and a good complement to the other existing algorithms \cite{DeLong:94b,Kane08}. As d-scan also offers access to both domains, it therefore appeared of interest to create a GP-based algorithm for d-scan.

\begin{algorithm}
\caption{Generalized projection algorithm for d-scan}\label{alg:GP}
\begin{algorithmic}[1]
\State $\phi^{(1)} \gets \rm{rand}(n)$ \Comment{initial guess phase}
\State $E^{(1)}$ from fundamental spectrum
\While{criteria not met}
\State Compute $\tilde{E}_{sig}(\omega,z)$ (Equation \ref{eq:signalfield})
\State $\tilde{E}_{\rm sig}'(\omega,z) \gets$ amplitude from d-scan  \Comment{data constraint}
\State $E_{\rm sig}'(t,z)$ from Fourier transform
\While{subcriteria not met}
\State Using NM, find $\phi''$ that minimizes $Z$ \Comment{physical constraint}
\EndWhile
\State $\tilde{E}_{\rm sig}''(\omega,z)$ from inverse Fourier transform
\State $\phi^{(1)} \gets \phi''^{(i)} $
\EndWhile
\State \textbf{return} $\varphi^{\rm(best)}$
\end{algorithmic}
\end{algorithm}

GP works by switching between two solution domains, and the transformation is accomplished by using a Fourier transform. Each solution domain satisfy a constraint. In FROG, these constraints are a data constraint, Eq.~(\ref{eq:dscantrace}), where agreement of the solution with the experimental FROG trace is enforced, and a physical constraint, Eq.~(\ref{eq:signalfield}), requiring that the solution follow the nonlinear process experimentally employed. The physical constraint is implemented in the time domain via the signal field $E_{\rm sig}(t,z)$. Starting with a single random seed for the phase $\phi$ and $|E|$ from the measured spectrum , one initially computes the signal field according to Eq.~(\ref{eq:signalfield}). The data constraint is enforced in the frequency domain by replacing the modulus of the signal by the square-root of the recorded d-scan trace:
\begin{equation} \label{eq:GP1}
\tilde{E'}_{\rm sig}(\omega,z) = \frac{\tilde{E}_{\rm sig}(\omega,z)}{\vert \tilde{E}_{\rm sig}(\omega,z) \vert} \sqrt{\left\lvert I_{\rm DS}^{(\rm (meas)}(\omega,z) \right\rvert}
\end{equation}
Subsequently, $\tilde{E'}_{\rm sig}(\omega,z)$ is transformed into the time domain by a Fourier transform. The physical constraint in \cite{DeLong:94b} is implemented by finding a phase $\phi''$ that gives rise to a signal field $E_{\rm sig}''(t,z)$ which, in turn, minimizes a distance metric
\begin{equation}
\label{eq:gpmetric}
Z = \sum^N_{i=1} \sum^M_{j=1} \left\lvert E''_{\rm sig}(t_i,z_j) -  E'_{\rm sig}(t_i,z_j) \right\rvert^2.
\end{equation}
After inverse Fourier transformation, the signal field $\tilde{E}_{\rm sig}''(\omega,z)$ then replaces $\tilde{E}_{\rm sig}(\omega,z)$ in Eq.~(\ref{eq:GP1}) in the next iteration of the algorithm. In \cite{DeLong:94b}, a one-dimension gradient-based minimization technique is used to find the best solution for the physical constraint. Given the integral form of the signal field Eq.~(\ref{eq:signalfield}), it does not appear straightforward to derive an analytical expression for the gradients. One could resort to a numerical evaluation of the gradients \cite{IFROG}, which would nevertheless substantially slow down the GP-based retrieval. We therefore chose to use NM for minimizing $Z$ instead. Our variant of GP for d-scan is summarized in Algorithm \ref{alg:GP}.

Figure \ref{fig:GP} shows the best retrieval results obtained with the GP algorithm without a penalty term. While the simple NM method only poorly reconstructs the spectral phase, GP comes very close, even though it cannot reproduce some fine detail in the original phase. However, as also shown in Table \ref{tab:algo}, the GP algorithm does not always reliably converge within the ~40 seconds that we allow for retrieval. In 3 out of 5 trials, the value of $G$ was actually larger than in simple NM. This problem can be overcome by allowing a significantly longer run time of the algorithm, see Fig.~\ref{fig:GP}. After more than an hour of processing time, the simple NM method is still stuck with very large oscillations in areas of low spectral power, also shown in Figure \ref{fig:GP}. Allowing for the same processing time, however, the GP algorithm reliably reconstructs the phase even in the extreme short-wavelength wing of the spectrum. We believe that the convergence can be further improved by one of the acceleration strategies successfully employed with FROG \cite{overcorrection,FROGBook}. In summary, we consider GP superior to NM as the general tendency of local stagnation is counteracted by a regular restarting mechanism for the Nelder-Mead sub-loop \cite{luersen2004globalized}. Consequently, there is less need to guess a suitable initial phase $\varphi$ with the GP algorithm, and there is an increased chance of observing a low d-scan error $G$ for repeated retrievals with different random seeds.

\section{Differential Evolution}

\begin{figure*}[htbp]
\centering
\includegraphics[width=\linewidth]{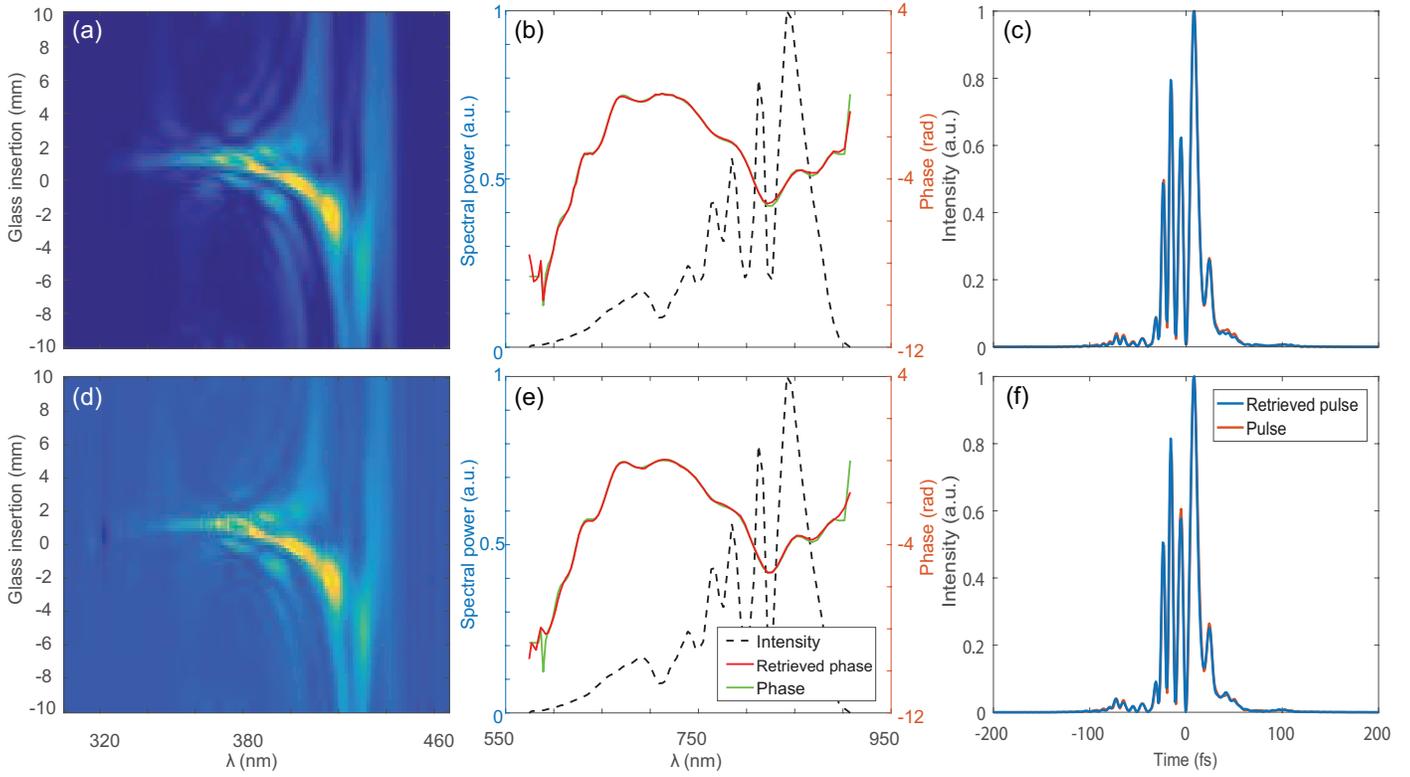}
\caption{Retrieved d-scan traces and spectral phases, using the Differential Evolution algorithm from traces with the minimum and maximum noise levels: (a-c) 0.1\%  and (d-f) 10\%, $G = 0.0879$.}
\label{fig:de}
\end{figure*}

\begin{figure*}[htbp]
\centering
\includegraphics[width=\linewidth]{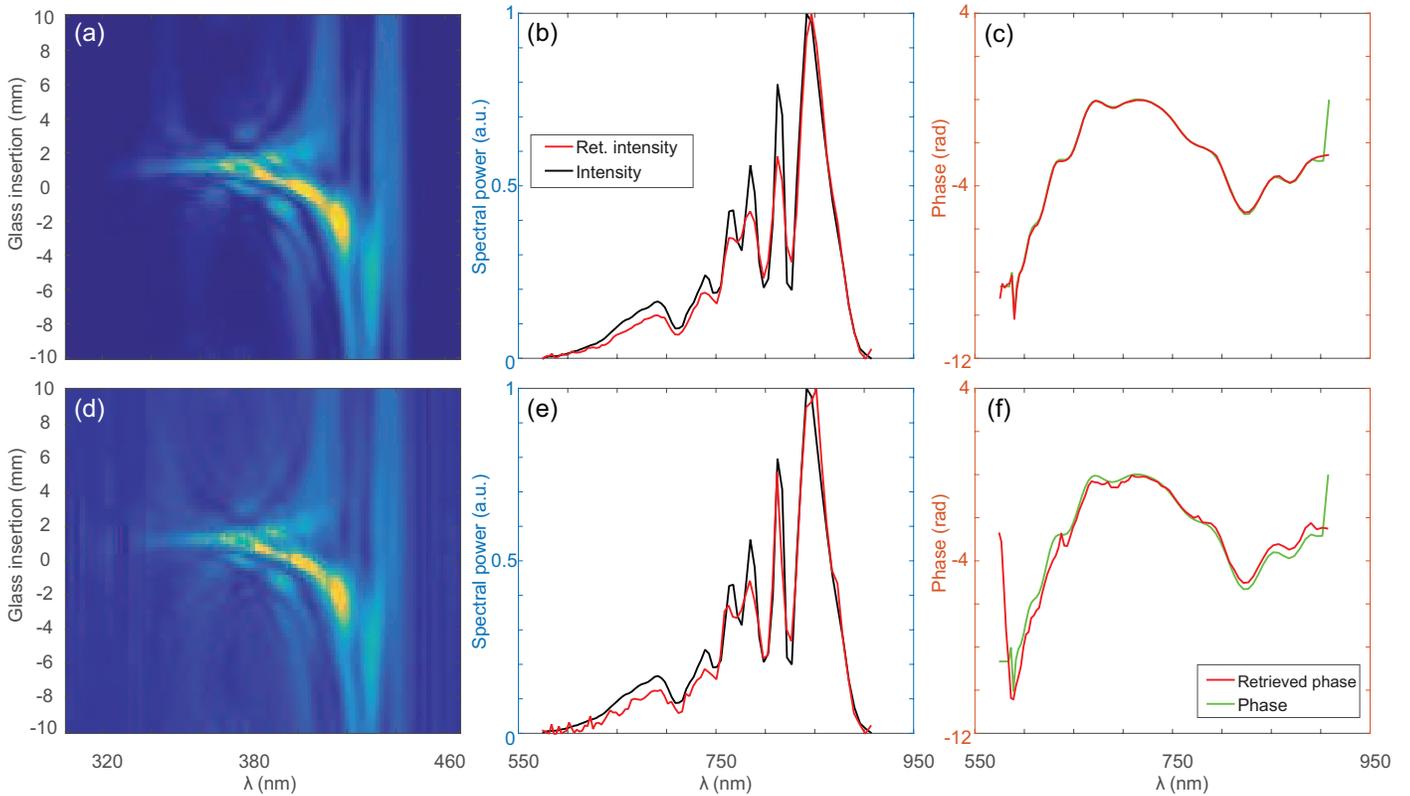}
\caption{Retrieval of both the spectral power and phase from the d-scan traces. (a-c) are retrievals from a trace with 0.1\% noise resulting to a $G = 0.0061$, while (d-f) are from a trace with 10\% noise, with $G = 0.0875$. The corresponding $\Delta \varphi_{\rm rms}$ are 0.073 and 0.329, with a computation time of two minutes each.}
\label{fig:both}
\end{figure*}

Evolutionary algorithms have been strongly inspired by nature, mimicking the biological selection process by mutating between a number of initial seeds $\varphi^{(i)}$ with the goal of determining the fittest subject in the population. Evolutionary algorithms have been quite popular for coherent control experiments in ultrafast spectroscopy, see, e.g., \cite{Gerber,Assion,Zeidler}. The evolutionary algorithms used in these experiments typically involve mutation and crossover, even though the detailed implementation varies. The potential usefulness of evolutionary strategies for pulse characterization has already been suggested in \cite{Gerber}, yet we are only aware of a few publications where evolutionary algorithms have been used for spectral phase retrieval from interferometric autocorrelations \cite{evolutionaryAC1,evolutionaryAC2}. A closely related genetic algorithm has been employed for pulse shape retrieval from FROG traces \cite{Nicholson:99}. We have evaluated a number of different evolutionary algorithms for phase retrieval from d-scan traces. Among these, we clearly found differential evolution (DE) superior, both concerning speed and convergence \cite{diffevolution,DEreview}. Compared to other evolutionary strategies, DE appears rather unique as it was specifically designed for optimizing continuous functions
that are internally represented by floating-point vectors, as opposed to arrays of bits.
Compared to other genetic strategies, DE uses similar operations such as selection,
mutation, and crossover between members, but the specifics and their order differ as detailed in Algorithm \ref{alg:DE}.

For the case of d-scan, an initial population with members $\phi^{(i)}$ is set up, with $i=1, \dots, D$, by sampling across the entire solution space using random numbers. The solution space can be limited to values from $-\pi$ to $\pi$ as phase is adequately represented by a wrapped function. A large population generally lessens the probability of stagnation at local minima, yet also slows down the algorithm. We found that a population of $D = 10$ members provides the best compromise for phase retrieval from the traces in Fig.~\ref{fig:dscantraces}. Nevertheless, this number may require adjustment depending on the complexity of the pulses under study. To evaluate the fitness of each member of the population, we use the same metric $G$ as in NM [Eq. \ref{eq:fitness}], with lower $G$ values indicating fitter population members. In the DE algorithm, some members undergo a mutation process, which can be implemented in various different ways \cite{Rocca:11}, which generally involves at least three members of the population. To this end, we employed a version of the best-of-random method \cite{lin2010synthesis}. For this method, one computes a mutant for each member $\phi^{(i)}$ of the population. Mutation is carried out with three randomly selected members of the remaining population, $\phi^{(a)}$, $\phi^{(b)}$, and $\phi^{(c)}$, which are arranged by decreasing fitness. The three are used to create a mutant according to
\begin{equation}
\label{eq:differential}
\phi^{(m)} = \phi^{(a)} + \mathrm{rand}(1) \left(\phi^{(b)}-\phi^{(c)}\right).
\end{equation}
The mutant then undergoes a crossover process with a primary parent $\phi^{(i)}$, and an offspring $\phi^{(o)}$ is then generated according to the following rule
\begin{equation}
\label{eq:crossover}
    \varphi_j^{(o)} =
\begin{cases}
    \varphi_j^{(m)} & \text{if } \mathrm{rand}(1) \leq \chi \\
    \varphi_j^{(i)}              & \text{else}
\end{cases}
\end{equation}
with $j=1, \dots ,n$ and $\chi$ is the crossover probability $\chi$. The crossover mechanism therefore mutates an adjustable fraction of the parent vector $\phi^{(i)}$. The process of mutation and crossover is repeated for all $\phi^{(i)}$ in the population, and then the fitness of all offsprings will be evaluated by Eq.~\ref{eq:fitness}. The parents and offsprings will be combined and sorted by fitness, and those whose ranks are beyond the population size $D$ will be discarded. This process is repeated generation after generation until one of the two criteria mentioned in Section \ref{sec:NM} is met. The whole algorithm is summarized in Algorithm \ref{alg:DE}. Using the notation for DE variants \cite{Storn1997}, this variant is DE/rand/1/bin, where the contributors to the mutant is randomly selected, and only one differential variation is used [Eq.~(\ref{eq:differential})] with a binomial crossover [Eq.~(\ref{eq:crossover})].

\begin{algorithm}
\caption{Differential evolution algorithm for d-scan}\label{alg:DE}
\begin{algorithmic}[1]
\State $\phi^{(i)} \gets 2\pi [\mathrm{rand}(n)] -\pi$ \Comment{initial population}
\State $f^{(i)} \gets G(\phi^{(i)})$ \Comment{fitness measurement}
\While{criteria not met}
\State sort($\phi^{(a)},\phi^{(b)},\phi^{(c)}$) (randomly generated)
\State $\phi^{(m)} \gets \phi^{(a)} + \mathrm{rand}(1)[\phi^{(b)}-\phi^{(c)}]$ \Comment{mutation}
\If{$r_j$ $\leq$ $\chi$} \Comment{crossover}
\State $\varphi_j^{(o)} \gets \varphi_j^{(m)}$
\Else
\State $\varphi_j^{(o)} \gets \varphi_j^{(i)}$
\EndIf
\State $f^{(o)} \gets G(\phi^{(o)})$ \Comment{fitness measurement}
\State sort($f^{(i, o)}$) and sort($\phi^{(i, o)}$)
\If{i, o $<$ D} \State discard $\phi^{(i, o)}, f^{(i,o )}$ \Comment{selection}
\EndIf
\EndWhile
\State \textbf{return} $\phi^{(\rm{best})}$
\end{algorithmic}
\end{algorithm}

Differential evolution can make large adjustments in every iteration since all phases act as parents, other than in the Nelder-Mead algorithm, where the adjustments are based most of the time only on the centroid. In addition, the changes in the population in DE are independent from each other, thus it is inherently parallel and can be programmed as such, which can further accelerate the algorithm. This extension to parallel computing has not been done yet in this manuscript, and remains a possibility for future work. The acceleration technique presented in \cite{Miranda:12b} can also be applied, wherein coarse representations of the spectral phase are used first prior to increasing the spectral resolution. Both of these techniques have been implemented for DE in this work. Unlike in NM, the increase in resolution is done at regular intervals, rather than waiting for the algorithm to stagnate first. This way is found to be helpful in speeding up DE, yet did not improve the NM algorithm.

The results of using the DE algorithm at various noise levels are shown in Figure \ref{fig:de}, and are also summarized in Table \ref{tab:de}. Even in the presence of fairly massive noise levels, the DE algorithm  correctly retrieves fine details in the spectral phase, even well into the wings of the spectrum. In the time domain [Figs.~\ref{fig:de}(c) and (d)], the relative intensities and durations of a fairly complicated sequence of 4 major and several minor sub-pulses are correctly reconstructed, with deviations on the few percent level, even in the presence of 10\% noise. It requires a similar noise level to significantly slow down convergence. Using optimized parameters ($\chi=0.5$, $D=10$), we typically observe convergence of the DE algorithm within less than 10 seconds on on a PC with 3.3\,GHz CPU frequency \cite{computer}, see Table~\ref{tab:de}.

A comparison of the different algorithms is shown in Table~\ref{tab:algo}. For the specific example case discussed in this manuscript, the rms deviation of the reconstructed phase and the target is below 100\,mrad for DE, which is to be compared to deviations exceeding one radian for all other techniques. While we occasionally observe sub-radian retrieval with the GP technique, the average performance of GP is not better than NM. In contrast, we never observed gross retrieval errors with DE and, additionally, DE converges about 4 times faster. It is worth noting again here that all the computation times for NM were recorded when the maximum number of function evaluations was reached, as opposed to DE, which always stopped with the convergence criterion. In addition to the individual techniques discussed so far, we also tried various combinations, e.g., of GP and DE as well as a combination of DE with regularization. The former combination lead to a performance that was in between the pure DE and GP algorithms. Only the combination of DE with regularization lead to a small acceleration of convergence together with an insignificant increase of the reconstruction error. While not discussed here in any detail, we also observe that accurate phase retrieval is accompanied by meaningful reconstruction of the spectral response function $R(\omega)$ of the $20\,\mu$m thick BBO crystal that was assumed in our simulations.

\begin{table}[htbp]
\centering
\caption{\bf Performance of DE at different noise levels averaged over 5 trials}
\begin{tabular}{cccc}
\hline
Noise Level & Comp Time (sec) & $\Delta \varphi_{\rm rms}$  & $G$ \\ \hline
0.1\%       & 9.4                 & 0.055   &  0.0033          \\
0.2\%       & 9.6                 & 0.054   & 0.0034             \\
0.5\%       & 9.8                 & 0.080   & 0.0058            \\
1\%         & 10.0                & 0.084   & 0.0107            \\
5\%         & 9.2                 & 0.066   & 0.0461            \\
10\%        & 21.1                & 0.122   & 0.0831            \\ \hline
\end{tabular}
\label{tab:de}
\end{table}

\begin{table}[htbp]
\centering
\caption{\bf Performance of various combinations of the techniques averaged over 5 trials}
\begin{tabular}{cccc}
\hline
Algorithm                & Comp Time (sec) & $\Delta \varphi_{\rm rms}$ & $G$ \\ \hline
NM                       & 40.5                & 2.36 & 0.0603                \\
DE                       & 9.4                 & 0.06 & 0.0033                \\
Regularized NM           & 39.4                & 2.30 & 0.0663                \\
Regularized DE           & 9.0                 & 0.07 & 0.0030                \\
GP$_{NM}$                & 43.2                & 4.25 & 0.0687                \\
GP$_{DE}$                & 17.7                & 1.80 & 0.0754                \\
Regularized GP$_{NM}$    & 44.2                & 2.38 & 0.0355                  \\
Regularized GP$_{DE}$    & 17.2                & 1.94 & 0.0747                \\ \hline
\end{tabular}
\label{tab:algo}
\end{table}

Due to the speed and reliability of DE, it is also possible to retrieve both the spectral power and the spectral phase at the same time, similar to what is usually done in FROG. To this end, we changed the algorithm such that each member of the population contains information both of the spectral amplitude and phase. In this form, the algorithm cannot simultaneously retrieve $R(\omega)$. Figure \ref{fig:both} shows the full retrieval results obtained with DE at two different noise levels. A slower increase of resolution is used due to the larger number of variables being optimized, and because of this, convergence now required 120.3 seconds for 0.1\% noise and 118.2 seconds for 10\% noise. Obtained values of $G$ are similar to what was observed in phase-only retrieval. Despite achieving good $G$ values, the resulting $\Delta \varphi_{\rm rms}$ is definitely larger, and it can be seen that the full-field retrieval is less robust towards noise compared to phase-only retrieval. This should not pose a major problem since the spectral power is usually easier to measure than the spectral response function.

\begin{figure}[htbp]
\centering
\includegraphics[width=\linewidth]{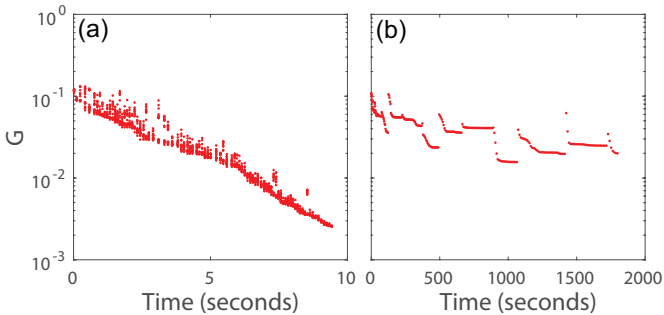}
\caption{Values of $G$ over time, for (a) DE and (b) NM, with same pulse used in the previous chapters.}
\label{fig:g}
\end{figure}

Another notable observation for DE is its strongly reduced tendency of getting stuck at local minima. This can be seen in Figure~\ref{fig:g}, where the DE-based retrieval resulted in a consistent reduction of the value of $G$ from the start, in contrast to NM where there were a lot of lock-ups throughout the retrieval.

\section{Analysis with various pulse shapes}

\begin{figure*}[htbp]
\centering
\includegraphics[width=.9\linewidth]{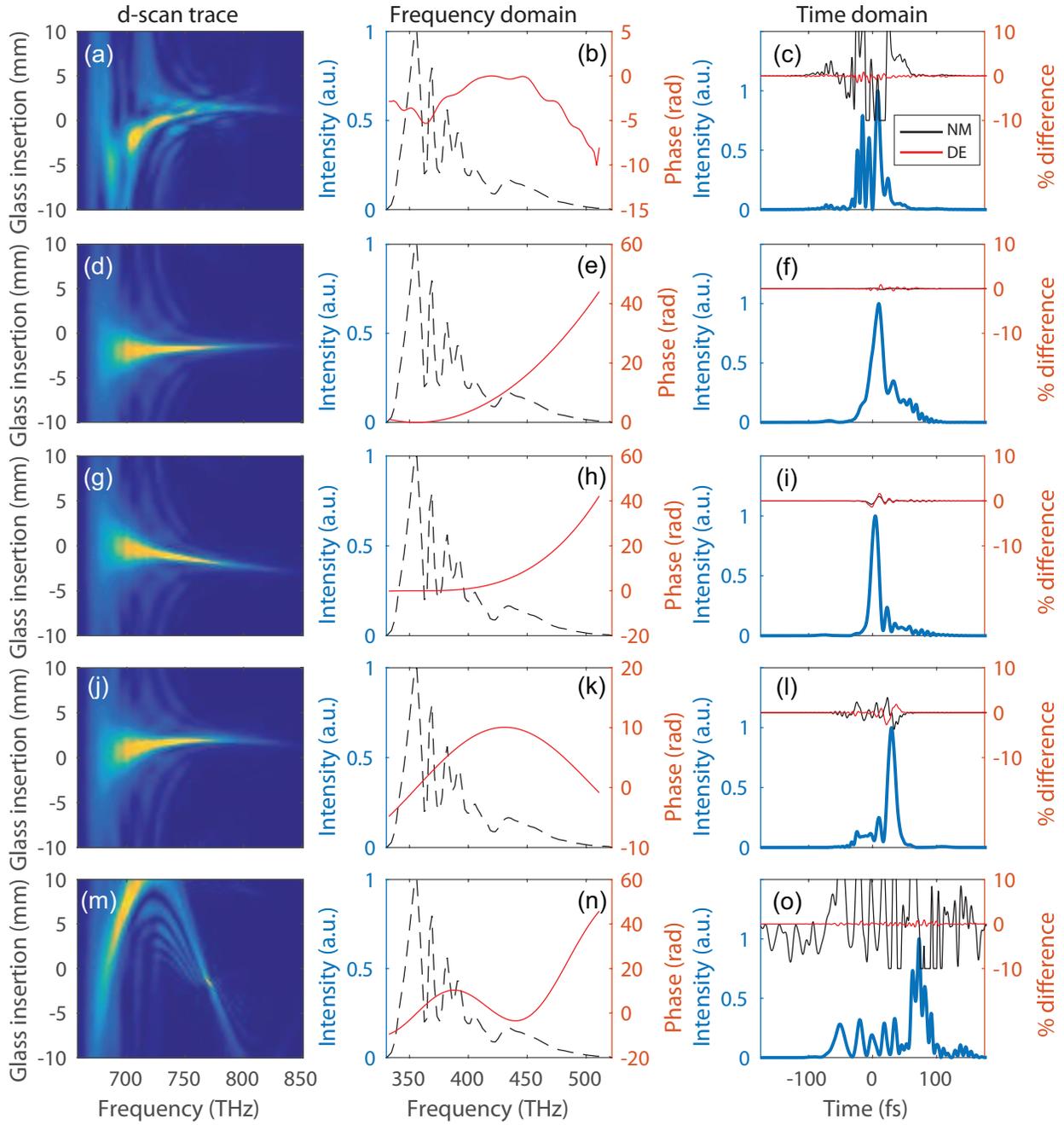}
\caption{D-scan traces and pulse structures in frequency and time domains for the different test cases. The differences of the pulse shapes for one of the NM and DE reconstructions are also shown in the righthand column. (a-c) ORIG, (d-f) GDD, (g-i) TOD, (j-l) TODR, and (m-o) GTR, according to the naming of the sets in Table~\ref{tab:pulses}.
Panels (c), (f), (i), (l), and (o) show the original temporal pulse profile in the bottom and the retrieval residual for DE and NM in the top. Residuals exceeding $\pm 10$\% were clipped.}
\label{fig:various}
\end{figure*}

To further evaluate the performance of the algorithms, five sets of pulses were used as test cases: (i) the original pulse from the previous chapters but represented with a smaller array, (ii) a set with various amounts of GDD, randomly generated between 0 and 100 fs$^2$, (iii) a set with various amounts of third-order dispersion (TOD) in the range between 0 to 300 fs$^3$, (iv) a set with the same range of TOD but with added phase ringing in the form of a sinusoid with amplitude of 10 radians and randomly selected frequency, and (v) a set similar to (iv) but with added GDD within the same range as in (ii). Thirty trials were done for each set, using the 8 retrieval variants in Table~\ref{tab:algo} and 6 noise levels in Table~\ref{tab:de}, resulting to a total of 7,200 recorded retrievals for this study. The original phase used previously is referred to as ORIG in Table~\ref{tab:pulses}, and the sets (ii)-(v) are referred to as GDD, TOD, TODR, and GTR, respectively. Figure~\ref{fig:various} features some of the pulses used in the study, along with the difference of the reconstructed pulses in time domain for a trial using NM and DE. The ringing of the phase functions in the last two rows are not obvious due to having small frequencies.

\begin{table}[ht]
\centering
\caption{Performance of NM and DE with different phase profiles}
\label{tab:pulses}
\begin{tabular}{ccccc}
\hline
                           & Algorithm      & Time (sec) & $\Delta \phi_{rms}$ & $G$      \\ \hline
\multirow{2}{*}{(i) ORIG}  & NM             & 24.6         & 3.7723              & 0.0729 \\
                           & DE             & 5.5          & 0.1331              & 0.0262 \\ \hline
\multirow{2}{*}{(ii) GDD}  & NM             & 23.7         & 0.0125              & 0.0268 \\
                           & DE             & 3.0          & 0.0436              & 0.0268 \\ \hline
\multirow{2}{*}{(ii) TOD}  & NM             & 23.8         & 0.0120              & 0.0269 \\
                           & DE             & 3.6          & 0.0542              & 0.0268 \\ \hline
\multirow{2}{*}{(iv) TODR} & NM             & 24.3         & 16.3328             & 0.0512 \\
                           & DE             & 4.1          & 0.2022              & 0.0264 \\ \hline
\multirow{2}{*}{(v) GTR}   & NM             & 24.3         & 20.8815             & 0.0475 \\
                           & DE             & 4.2          & 0.3076              & 0.0275 \\ \hline
\end{tabular}
\end{table}

In all these retrievals, DE outperformed NM in terms of speed. Moreover, NM also often failed to converge within the $\approx 25$\,s that we allowed for retrieval. For cases (ii) and (iii) with their rather simple spectral phases, NM converges throughout. In this situation, the resulting $G$ values are comparable to DE, yet the rms phase errors are lower than those obtained with DE. Nevertheless, these lower rms values appear to be the only advantage that we could observe for the NM technique. In the time domain, we observe that DE converges to the original pulse shape with peak deviations of 2\%. The performance of DE appears rather independent of spectral phase. When NM converges it delivers even better reconstructions. However, when it fails, deviations are in the range of many ten percent.

\section{Conclusions}

From the above observations, we conclude that d-scan is well suited for phase-only retrieval, yet appears less suitable for a full reconstruction of the pulse shape without an independently measured spectrum. Moreover, a FROG-like marginal check appears less dependable for d-scan traces, and the retrieved response functions $R(\omega)$ should be discussed with some caution. Nevertheless, the demonstrated insensitivity of d-scan towards the exact spectral power can also be considered an advantage of the method. In fact, d-scan appears to be much more selectively sensitive to the phase structure of a pulse, and this fact appears completely understandable from the structure of the signal field in d-scan.

With the DE based retrieval method, d-scan enables precise measurements of the spectral phase structure of complicated supercontinuum pulses as they are generated in hollow-fiber compressors \cite{Nisoli} or in filament compression experiments \cite{Stibenz}. Considering that d-scan is a self-referenced technique, it shows remarkable capabilities in accurately reconstructing even fairly complicated pulse shapes consisting of several sub-pulses. These capabilities are only marginally affected by the presence of noise.
Even in the absence of dynamical instabilities \cite{Michelle}, the supercontinuum scenario has proven as a serious test for any given pulse characterization technique \cite{Balciunas}. A particular problem is the appearance of pronounced spectral dips, which may cause phase artifacts in SPIDER \cite{HDRSPIDER}. We also found it difficult and time-consuming to correctly retrieve the exact spectral phase and pedestal structure of complex pulses with self-referenced FROG using generalized projections \cite{Xu}. As DE is very flexible, however, the discussed algorithm can be easily adapted to the signal fields encountered in FROG. While generalized projections can be very efficiently coded for FROG signal fields, we do not expect that DE will outperform existing retrieval software in terms of convergence speed. Nevertheless, we are confident that DE may help to overcome the problem of local stagnation in generalized-projections based FROG retrieval. Moreover, DE is much less susceptible to experimental noise than any of the other techniques under test. In particular, DE may also prove useful for phase-only retrieval of FROG traces with complex structure. We therefore conclude that differential evolution may substantially enhance the toolbox of ultrafast pulse characterization techniques.

\section*{Funding Information}
Deutsche Forschungsgemeinschaft (DFG) (STE 762/11-1)
 
\section*{Acknowledgments}
We acknowledge fruitful discussions with Rick Trebino (Georgiatech) and Bernd Hofmann (TU Chemnitz). We thank Mathias Hoffmann, Bruno Chanteau, Christoph Jusko, and Uwe Morgner at Leibniz-University Hannover for their contributions to an early version of the Nelder-Mead code.

\end{document}